
\magnification=1200
\baselineskip=18pt
\centerline{\bf
Meson phase space density from interferometry}
\centerline{
George F. Bertsch}
\centerline{Dept. of Physics and Institute for Nuclear
Theory, FM-15}
\centerline{University of Washington}
\centerline{Seattle, WA 98195}

\noindent
Abstract

\noindent
The interferometric analysis of meson correlations provides
a measure
of the
average phase space density of the mesons in the final
state. This quantity is a useful indicator of the statistical
properties of the system,
and it can be extracted with a minimum of model assumptions.  Values
obtained from recent measurements are consistent with the
thermal value, but do not rule out superradiance
effects.
\vskip 1.3cm
It would be interesting to know the average phase space density of the
pions produced in ultrarelativistic heavy ion collisions.  In the
final state, the local phase space density is frozen (Liouville's
theorem) and it gives a measure of the dynamics in the prior
interacting region.  If the system could be described by a local
statistical equilibrium, the distribution function $f$ would
have the
Bose-Einstein form, $ f_T(p,r) = 1/(\exp(u(r)\cdot p/T) -1) $.
For massless particles the
average of this quantity is a pure number,
$$ \langle f \rangle_T = {\int d^3pd^3r f_T^2 \over \int d^3pd^3r
f_T}={\zeta(2)\over\zeta(3)}-1\approx
0.37\,. $$
For massive bosons, the number is lower; for example, for
pions at a chemical freezeout temperature of 200 MeV the
average is $$
\langle f\rangle_{m,T} \approx 0.15\,. \eqno(1)$$
If the experimental $\langle f\rangle $ were close to this, it would be
welcome evidence for the existence of a local equilibrium.  If
$\langle f\rangle $ came out much larger, it would lend
considerable support to the idea of superradiant pion
states[1], which have been an object of renewed
interest[2,3].
On the other hand, if
$\langle f\rangle $ came out much smaller than thermal, it would
point to entropy-generating processes such as the slow decay of
heavy resonances or quark-gluon droplets.

In this letter I want to point out that
the measured two-particle correlations
yield direct information about the phase space density of the mesons,
when interpreted according to Pratt's interferometric
formula[4].
To see this, we begin from Pratt's formula written as

$${d^6 n^{(2)} \over d^3 p_1 d^3 p_2}\Big|_{p+q,p-q} -
{d^3 n^{(1)} \over d^3p}\Big|_{p+q}
{d^3 n^{(1)} \over d^3p}\Big|_{p-q} =
\int d^4x_1d^4x_2 g(x_1,\vec{p})g(x_2,\vec{p}) \,\cos q\cdot (x_1-x_2)
\eqno(2)$$
where $g$ is the source function for the mesons.  We next
convert the source function $g$ to an equivalent source at
a common time $t_0$ by the replacement
$$ g(\vec{r},t,\vec{p})\rightarrow
\delta(t-t_0) \int^{t_0} dt'
g(\vec{r}-\vec{v}(t'-t_0),t',\vec{p})\equiv
\delta(t-t_0) (2\pi)^3{f(r,p)} $$
where $v$ is the velocity associated with the momentum
vector $\vec{p}$.
This replacement does not affect the correlation function if
$\vec{v} \cdot (\vec{p_1}-\vec{p_2}) = (E_1-E_2)$. The condition is
satisfied for
small momentum differences and for longitudinal motion of
extreme relativistic particles, and seems rather safe for
the
present application.  The coefficient of the
$\delta$-function is then the phase space density at $t_0$,
extrapolating the positions of the final state mesons to that
time.  We next integrate over the momentum difference $d^3q$,
which produces a delta function $\delta^3(r_1-r_2)$ to
eliminate one of the spatial integrals.  The result is
$$\int {d^3q\over (2\pi)^3}\int d^4x_1d^4x_2
g(x_1,\vec{p})g(x_2,\vec{p})
\,\cos q\cdot (x_1-x_2)
= (2\pi)^3\int d^3r f^2(\vec{r},\vec{p})\,.
$$
Finally we integrate over $d^3p/(2\pi)^3$ and normalize to the number of
particles to obtain the phase space average,
$$\langle f\rangle  = {1 \over n} \int d^3 p \int d^3 q \Bigg[
{d^6 n^{(2)} \over d^3 p_1 d^3 p_2}\Big|_{p+q,p-q} -
{d^3 n^{(1)} \over dp^3}\Big|_{p+q}
{d^3 n^{(1)} \over dp^3}\Big|_{p-q} \Bigg]\,. \eqno(3)
$$
For heavy ion collisions it is more useful to make the average
over a fixed rapidity interval, because the system evolves
to produce a
spatial separation between particles of different
rapidities.  The different rapidity groups equilibrate
independently.
The formula
for a small rapidity interval reads
$$ \langle f \rangle_{dy} = {1\over dn/dy} \int {d^2p_t\over p_0}
\int d^3 q \Bigg[
{d^6 n^{(2)} \over d^2 p_{t1}dy_1 d^2 p_{t2}dy_2}\Big|_{p+q,p-q} -
{d^3 n^{(1)} \over d^2p_tdy}\Big|_{p+q}
{d^3 n^{(1)} \over d^2p_tdy}\Big|_{p-q} \Bigg]\,.
\eqno(4)$$
Eq. (3) and (4) just require integral properties of the
correlation function, so they should be less dependent on
the accuracy of the momentum measurements than other
observables.

The integral should undoubtedly be evaluated directly
from the experimental data, but for an orientation I shall
try to evaluate it from published Na44 parameterized
distributions[5]. The Na35 experiment[6] obtained similar
information, but did not quote the entire parameterization.
One of the common parameterizations for these correlations
is Gaussian with parameters $\lambda$ and source sizes
$R_L,R_s$ and $R_o$,
$$
{d^6 n^{(2)} \over d^2 p_{t1}dy_1 d^2 p_{t2}dy_2}
= \bigg(1+\lambda\exp(-{1\over 2}(q_L^2R_L^2+q^2_sR_s^2+q_o^2R_o^2)
\bigg)
{d^3 n^{(1)} \over d^2p_tdy}\Big|_{p_1}
{d^3 n^{(1)} \over d^2p_tdy}\Big|_{p_2}\,.
$$
The single-particle transverse momentum spectra can be
parameterized by an exponential function,
$$
{d^3 n^{(1)} \over d^2p_tdy} =
{dn \over dy} {exp(-p_t/T_t)\over 2\pi T_t^2}\,.
$$
With this parameterization, the average phase space density
is given by
$$
\langle f \rangle_dy = \sqrt{\pi\over 2} {1\over
R_LR_sR_oT_t^3}\,.
$$
Ref. [5] quotes the following numbers for S + Pb$\rightarrow
\pi^+ +$X at midrapidity:  $\lambda \approx 0.4$, $R_t\approx6.0$ fm, $R_L
\approx6.0$ fm and
$dn/dy\approx 40 $.
{}From their Fig. 7 can be deduced $T_t
\approx 187 $ MeV/c. Taking $R_s=R_o=R_t$ and
inserting these numbers in eq. (3), I
obtain
$$  \langle f \rangle_{dy} \approx 0.07-0.16\,. $$
The range is obtained from the quoted experimental errors
combined quadratically.  I note that the NA35 experiment
found a somewhat smaller source size;
the lack of agreement
between experiments is a caution not to draw strong
conclusions from the present data.

One should also be reminded that the assumptions going into
the interferometric formula, eq. (2), may not be well
satisfied.  The most critical assumption is that the
one-particle distribution function is not affected by
the Bose symmetrization, which is only satisfied for
low phase space densities.

Given these caveats, what does one conclude?  The extracted
phase space density is consistent with local
statistical equilibrium for a chemical freezeout temperature in
the 100-200 MeV range.  The analysis includes pions from
long-lived resonance decays such as $\omega\rightarrow3\pi$
and $\eta\rightarrow 3 \pi$.  If these could be subtracted
out, the phase space density would be somewhat higher.  Thus
there may be an excess of pions produced in the source, and
the possibility of coherent pion effects should not be ruled
out.

The extracted phase space density appears high enough to
make unlikely that long-lived intermediates
such as droplets of quark-gluon plasma are produced
abundantly. This is
consistent with the present theoretical expectation of
no strong first-order phase transition in the quark-gluon
plasma[7].

I wish to thank M. Herrmann, H. Gutbrodt, J. Kapusta and E.
Shuryak for discussions.  This work carried out at the
Institute for Theoretical Physics and is supported partly by
the Department of Energy under grant DE-FG06-90ER40561 and
partly by the National Science Foundation under Grant PHY
89-04035.

\noindent
{\bf References}
\item{[1]}  C.S. Lam and S.Y. Lo, Phys. Rev. Lett. 52 (1984)
1184.
\item{[2]} S. Pratt, Phys. Lett. B301 (1993) 159.
\item{[3]} K. Rajogopal and F. Wilczek, Nucl. Phys. B399
(1993) 395.
\item{[4]}  S. Pratt, Phys. Rev. Lett. 53 (1984) 1219.
\item{[5]}  H. Atherton, et al., Nucl. Phys. A544 (1992) 125c.
\item{[6]}  J. Baechler, et al., Nucl. Phys. A544 (1992)
293c.
\item{[7]}  F.R. Brown, F.P. Butler, H. Chen, et al., Phys.
Rev. Lett. 65 (1990) 2491.
\bye